\documentclass[aps,preprint,showpacs,preprintnumbers,amsmath,amssymb,floatfix]{revtex4}
\usepackage{graphicx}
\begin{document}
\def\brho {\mbox{\boldmath $\rho$}}
\def\r {{\bf r}}

\def\beq{\begin{equation}}
\def\eeq{\end{equation}}
\def\beqn{\begin{eqnarray}}
\def\eeqn{\end{eqnarray}}
\def\souligne#1{$\underline{\smash{ \hbox{#1}}}$}

\title{On-top fragmentation stabilizes atom-rich attractive Bose-Einstein condensates}

\author{Lorenz S. Cederbaum, Alexej I. Streltsov, and Ofir E. Alon}

\affiliation{Theoretische Chemie, Physikalisch-Chemisches Institut, Universit\"at Heidelberg,\\
Im Neuenheimer Feld 229, D-69120 Heidelberg, Germany}

\begin{abstract}
It is well known that attractive condensates do not posses a stable ground state in 
three dimensions.
The widely used Gross-Pitaevskii theory predicts the existence
of metastable states up to some critical number $N_{\mathrm{cr}}^{\mathrm{GP}}$ of atoms.
It is demonstrated here that fragmented metastable
states exist for atom numbers well above $N_{\mathrm{cr}}^{\mathrm{GP}}$.
The fragments are strongly overlapping in space.
The results are obtained and analyzed analytically as well as numerically.
The implications are discussed.
\end{abstract}

\maketitle

Attractive Bose-Einstein condensates (BECs)
have attracted much experimental \cite{e1,e2,e3,e4,e5,e6}
and theoretical \cite{t1,ga1,t2,ct1,ga2,t4,ga3,t5,ct2,t6,t7,t8,t9,t10,Pethick,Pit_book} interest.
The finding that the interaction between the atoms of a BEC
can be varied systematically \cite{e2},
allows one to prepare attractive condensates out of repulsive ones and to control them.
A unique feature of attractive condensates in three dimensions is that they collapse,
i.e., they do not posses a stable ground state as repulsive condensates do.
At a given interaction strength,
an attractive BEC can be formed, however, in a metastable
state if the number of its atoms is smaller 
than some critical number $N_{\mathrm{cr}}$.

The non-linear Schr\"odinger equation, also called Gross-Pitaevskii (GP) equation,
is widely used to study condensates \cite{Pethick,Pit_book}.
This equation naturally predicts the collapse of attractive condensates
and the appearance of metastable states, see, e.g.,
Refs.~\cite{t1,ga1,t2,ct1,ga2,ga3,ct2,Pethick,Pit_book}.
The GP result can be derived using mean-field theory.
Assuming a many-body wavefunction 
$\Psi_{\mathrm{GP}}(\vec r_1,\vec r_2,\ldots,\vec r_N)=\phi_0(\vec r_1)\phi_0(\vec r_2)\ldots\phi_0(\vec r_N)$,
which possesses the structure of the wavefunction of $N$ noninteracting indistinguishable
bosons, one readily determines the GP equation by minimizing the energy expectation value
$E_{\mathrm{GP}}=\left<\Psi_{\mathrm{GP}}\left|H\right|\Psi_{\mathrm{GP}}\right>$
with respect to the orbital $\phi_0(\vec r)$.
Using, as usual, the Hamiltonian $H=H_0+\hat W$, where $H_0=\sum_i^N \hat h(i)$
describes $N$ bosons in an external trap and $\hat W$ the contact interaction between the bosons,
the GP energy reads \cite{Pethick,Pit_book}
\beq\label{e1}
 E_{\mathrm{GP}} = Nh_{00} + \lambda_0 \frac{N(N-1)}{2} \int \left|\phi_0(\vec r)\right|^4 d^3r, 
\eeq
where $h_{00}=\big<\phi_0\big|\hat h\big|\phi_0\big>$
and the interaction strength $\lambda_0$ is proportional to the s-wave scattering length and
takes on negative values for attractive condensates.

\begin{figure}
\includegraphics[width=12cm, angle=-0]{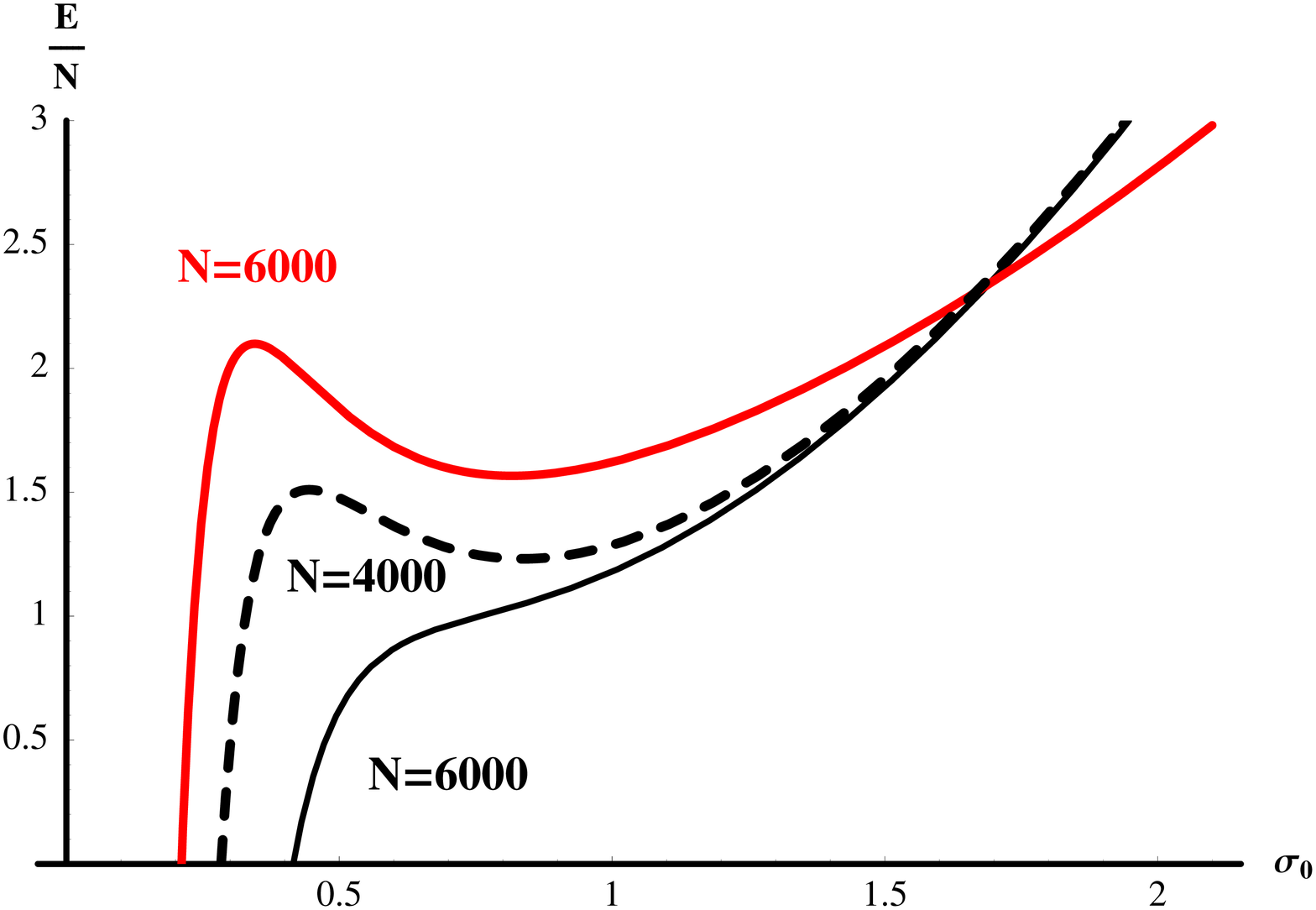}
\caption {(Color online) Energies per particle
$E/N$ as a function of the scaling parameter $\sigma_0$.
Shown are the GP energies for 
$N<N_{\mathrm{cr}}^{\mathrm{GP}}$ and $N>N_{\mathrm{cr}}^{\mathrm{GP}}$ (in black)
and the energy of the fragmented condensate  for
$N>N_{\mathrm{cr}}^{\mathrm{GP}}$ (in red); see text for more details.
$N$ is indicated on the respective curves. 
The relative fragmentation is $\frac{n_1}{N}=0.15$.
The scenario analyzed corresponds to $^{85}$Rb atoms in an isotropic harmonic trap of 
frequency $\omega=2\pi\cdot 15$Hz and s-wave scattering length
of $a_s=-7.14a_0$, where $a_0$ is Bohr radius, which are experimentally accessible \cite{e2}. 
For these parameters $N_{\mathrm{cr}}^{\mathrm{GP}}=5000$.
The energy is expressed in units of $\hbar\omega$ and length 
in units of $\frac{1}{\beta}=\sqrt{\frac{\hbar}{{\mathrm m}\omega}}$. 
Transforming to dimensionless units in which $\hbar={\mathrm m}=\omega=1$,
the corresponding interaction strength is $\lambda_0=4\pi a_s\beta=-0.001684$.
}
\label{f1}
\end{figure}

The collapse and the existence of metastable states has been nicely discussed 
in the framework of the time-independent \cite{Pethick,ga1,ga2,ga3}
as well as time-dependent \cite{ct1,ct2} GP equation 
which give identical results.
Here, we follow the time-independent formulation.
We consider without lost of generality a spherical harmonic trap and,
as is common to all these approaches,
we assume a normalized radial Gaussian function
$\varphi_0(r)=(\frac{\beta}{\sigma})^{3/2}\frac{2}{\pi^{1/4}}e^{-\frac{1}{2}(\beta r/\sigma)^2}$
for the orbital $\phi_0(\vec r)= (\frac{1}{4\pi})^{1/2}\varphi_0(r)$
with $\beta=(\frac{{\mathrm m}\omega}{\hbar})^{1/2}$.
For $\sigma=1$, $\phi_0$ is the eigenfunction of a particle of mass ${\mathrm m}$
in the harmonic trap of frequency $\omega$.
Inserting this orbital in (\ref{e1}) one readily obtains 
\beq\label{e2}
  E_{\mathrm{GP}}/\left(\frac{3}{4}\hbar\omega N\right) = \sigma^{-2}+\sigma^2 + \Lambda_0\frac{N-1}{2}\sigma^{-3},
\eeq
where $\Lambda_0=\frac{\lambda_0}{4\pi}(\frac{2}{\pi})^{1/2}\beta^{3/2}\frac{4}{3\hbar\omega}$.
Since $\Lambda_0$ is negative for an attractive BEC,
this energy is unbounded from below.
For $\sigma \to 0$,
the kinetic energy $\sigma^{-2}$ grows, but cannot compensate the negative
interaction term which scales as $\sigma^{-3}$, and
the condensate collapses such that its density approaches a $\delta$-function.
$E_{\mathrm{GP}}(\sigma)$ exhibits a local minimum for $N<N_{\mathrm{cr}}^{\mathrm{GP}}$
as is illustrated in Fig.~\ref{f1}. 
To find the critical atom number $N_{\mathrm{cr}}^{\mathrm{GP}}$,
we set to zero the first and the second derivative of $E_{\mathrm{GP}}(\sigma)$ 
with respect to $\sigma$, and obtain \cite{Pethick,ga1,ga2,ga3}
\beq\label{e3}
N_{\mathrm{cr}}^{\mathrm{GP}} = \left(\frac{1}{5}\right)^{1/4}\frac{16}{15}\frac{1}{|\Lambda_0|}, \qquad
\sigma_{\mathrm{cr}}=\left(\frac{1}{5}\right)^{1/4}
\eeq
in agreement with the result found using time-dependent approaches \cite{ct1,ct2}.
The latter approaches also show that for $\sigma=r$ the r.h.s. of (\ref{e2}), i.e.,
$E_{\mathrm{GP}}/N$, plays the role of an effective potential for the motion
of an atom at position $r$ in the condensate.

Within the GP theory, BECs cannot posses metastable states for $N>N_{\mathrm{cr}}^{\mathrm{GP}}$.
The GP theory is a single-orbital mean-field theory in
which all bosons reside in a single orbital $\phi_0(\vec r)$.
We shall show below that within the multi-orbital mean-field theory \cite{PLA03,PLA05},
which contains the GP theory as a special case,
BECs do posses metastable states for $N$ substantially larger than $N_{\mathrm{cr}}^{\mathrm{GP}}$.
In this theory the many-body wavefunction is 
$\Psi=\hat{\cal S}\phi_0(\vec r_1)\ldots\phi_0(\vec r_{n_0})\phi_1(\vec r_{n_0+1})
\ldots\phi_1(\vec r_{n_0+n_1})\ldots$
where $\hat{\cal S}$ is the symmetrizing operator,
and describes the situation in which the $N$ bosons are distributed among several orbitals:
$n_0$ bosons reside in an orbital $\phi_0(\vec r)$,
$n_1$ bosons in $\phi_1(\vec r)$,
and so on.
Let us for transparency return to spherical traps
and express the orbitals as
$\phi_{lm}(\vec r)=\varphi_l(r)Y_{lm}(\theta,\varphi)$.
We consider the situation where the density $\rho(\vec r)$
is spherically symmetric, as is the case also for GP.
Consequently, the occupation numbers $n_{lm}$ corresponding to the $\phi_{lm}$
do not depend on $m$ and may be denoted $n_l$.
The energy expectation value
$E=\left<\Psi\left|H\right|\Psi\right>$ now reads
\beqn\label{e4}
& & E=n_0h_{00} + 3n_1h_{11} +
  \frac{\lambda_0}{4\pi} \Bigg\{\frac{n_0(n_0-1)}{2} \int_0^\infty\left|\varphi_0(r)\right|^4 r^2dr+ \nonumber \\ 
 & & \left[\frac{69}{10} n_1^2 + \frac{21}{10} n_1 \right]
\int_0^\infty\left|\varphi_1(r)\right|^4 r^2dr +
6n_0n_1\int_0^\infty\left|\varphi_0(r)\right|^2\left|\varphi_1(r)\right|^2 r^2dr \Bigg\} \
\eeqn
where for simplicity we have considered the four orbitals
spanned by $l=0,m=0$ and $l=1,m=0,\pm 1$.
Of course, $N=n_0+3n_1$.
It is easily seen that this energy reduces to the GP energy 
$E_{\mathrm{GP}}$ in (\ref{e1}) if we put $n_1=0$.

The many-body states described by the multi-orbital mean-field
theory are fragmented states.
In a fragmented state more than one eigenstate of the one-particle density
operator is macroscopically occupied \cite{frag0}.
In the GP theory the density reads $\rho_{\mathrm{GP}}(\vec r)=N|\varphi_0(r)|^2/(4\pi)$
and as all atoms reside in a single orbital, this theory is 
unable to describe fragmented states.
In the multi-orbital theory the density 
takes on the appearance
\beq\label{e5}
 \rho(\vec r)=\sum_{l,m} n_l|\phi_{lm}(\vec r)|^2 = \sum_{l=0} n_l(2l+1)|\varphi_l(r)|^2/(4\pi)
\eeq
and the $n_l$ describe the macroscopic occupations of the fragments.

Fragmentation is by now well established in repulsive condensates \cite{Sipe,ALN,path,Erich}.
There, as the repulsion between the atoms increases,
the atoms minimize the energy by building fragments which avoid each other.
At sufficiently strong interaction,
each atom may even reside in its own
orbital in analogy to fermions \cite{path}.
In the limit of infinite repulsion this behavior is known
as the Tonks-Girardeau limit \cite{Girardeau}.
In principle, even the famous Mott-insulator
state of $N$ atoms in $N$ wells of an optical lattice \cite{Fisher,Zoller,Bloch}
can be viewed as a fragmented state,
as each atom resides in its own orbital strongly localized in a different well.
Indeed, the transition from superfluid to Mott-insulator states
can be well understood within the multi-orbital mean-field theory \cite{z0}.
In contrast to repulsive systems,
fragmentation is rather counter intuitive in attractive condensates
as the atoms attract each other and would like
to be close to each other.
There are, however, first hints in the literature
for fragmentation in attractive systems,
the examples being atoms in a ring \cite{Erich,Ueda97,Ueda03,EPL} 
and in a symmetric one-dimensional double-well potential \cite{PLA03}.
Interestingly, the fragmentation in these examples
is dictated by symmetry and disappears when the symmetry of the
trap is slightly distorted \cite{Ueda97,PLA03}.
The fragmentation in three dimensions
discussed in the present work is not dictated by symmetry.
We have performed the analogous computations for an anisotropic trap and came to the same
conclusions as for the isotropic ones.

We now return to (\ref{e4}) and compute the relevant quantities for the harmonic oscillator.
Each radial fragment orbital, in principle,
contracts and collapses differently,
and we give each $\varphi_l(r)$ a different scaling $\sigma_l$.
The $l=0$ orbital is chosen as for the GP equation (with $\sigma=\sigma_0$)
and for $l=1$ we use the common respective radial harmonic
$\varphi_1(r)=\sqrt{\frac{8}{3}}\frac{1}{\pi^{1/4}}(\frac{\beta}{\sigma_1})^{3/2}(\frac{\beta r}{\sigma_1})
e^{-\frac{1}{2}(\beta r/\sigma_1)^2}$.
With this natural choice the energy expectation value now takes on the following explicit form
\beqn\label{e6}
 & & E(\sigma_0,\sigma_1)/\left(\frac{3}{4}\hbar\omega\right) =
 n_0\left(\sigma_0^{-2}+\sigma_0^2\right) + 5n_1\left(\sigma_1^{-2}+\sigma_1^2\right) + \nonumber \\
 & & + \Lambda_0\left\{
 \frac{n_0(n_0-1)}{2}\sigma_0^{-3} + \left[\frac{69}{10} n_1^2 - 
\frac{21}{10} n_1\right]\frac{5}{12}\sigma_1^{-3} + 
3n_0n_1 \frac{2^{5/2}\sigma_0^2}{\left(\sigma_0^2+\sigma_1^2\right)^{5/2}}
\right\}. \
\eeqn
If one divides $E$ by $N$,
one can immediately express the result for large $N$ in terms of $\sigma_0$, $\sigma_1$, $\Lambda_0N$
and the relative occupancy $\frac{n_1}{N}$ only.

At a given relative occupancy it is easy to compute the optimal values of $\sigma_0$ and $\sigma_1$ 
minimizing the energy $E$ in (\ref{e6}).
An illustrative result for a cut along $\sigma_0$ of the energy $E$ through
its minimum is depicted in Fig.~\ref{f1}
for $N>N_{\mathrm{cr}}^{\mathrm{GP}}$ and compared to the GP energy
$E_{\mathrm{GP}}$ for the same $\Lambda_0$ and $N$.
It is clearly seen that while GP theory predicts the collapse of the condensate,
fragmentation does lead to the appearance of a metastable state.
In addition, one notices that $E_{\mathrm{GP}}$ is larger than $E$ in (\ref{e6}) for
a whole range of $\sigma_0$ (see Fig.~\ref{f1}).
Interpreting $E(\sigma_0=r)$ as an effective potential as done, e.g., 
in \cite{ga1,ct1,ga2,ga3,ct2} for $E_{\mathrm{GP}}$,
this may have relevant implications also on the dynamics of an attractive condensate 
initially prepared as a repulsive condensate which naturally has a broad distribution 
(i.e., a large $\sigma_0$)
and then switching the interaction to negative values.
The thus created attractive condensate is likely to fragment as its momentarily 
lowest energy state is a fragmented one.

\begin{figure}[ht]
\includegraphics[width=12cm,angle=-0]{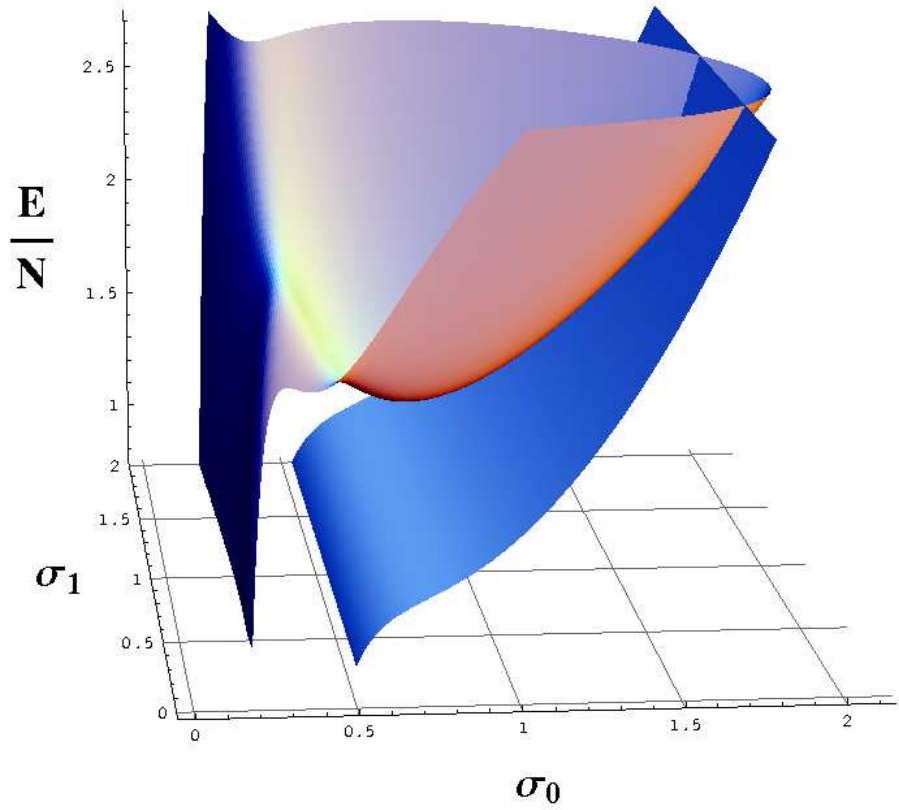}
\caption [kdv]{(Color online) The energy per particle 
$E(\sigma_0,\sigma_1)N$
as a function of $\sigma_0$ and $\sigma_1$
(see Eq.~(\ref{e6}))
compared to the respective GP energy
(see Eq.~(\ref{e2})). $N=6000$.
The quantities used are the same as in Fig.~\ref{f1}.
For these quantities the minimum of $E$ is at
$\sigma_0=0.816$ and $\sigma_1=0.811$.
}
\label{f2}
\end{figure}

The energy $E$ is shown in Fig.~\ref{f2} as a function of both
$\sigma_0$ and $\sigma_1$.
This function exhibits a clear minimum.
It is also seen that this minimum refers to a
metastable state as $E$ is unbounded from below for 
$\sigma_0,\sigma_1 \to 0$.
Numerically, it is straightforward to determine the critical number of
atoms at which $E(\sigma_0,\sigma_1)$ in (\ref{e6}) ceases to support metastable states for a given fragmentation.
In Fig.~\ref{e3} this critical number $N_{\mathrm{cr}}$ is drawn 
as a function of the relative occupancy $\frac{n_1}{N}$.
Since $N=n_0+3n_1$, the value of $\frac{n_1}{N}$ can maximally be $\frac{1}{3}$.
It is clearly seen in the figure that $N_{\mathrm{cr}}$ can be much large than 
$N_{\mathrm{cr}}^{\mathrm{GP}}$.

\begin{figure}[ht]
\includegraphics[width=12cm,angle=-0]{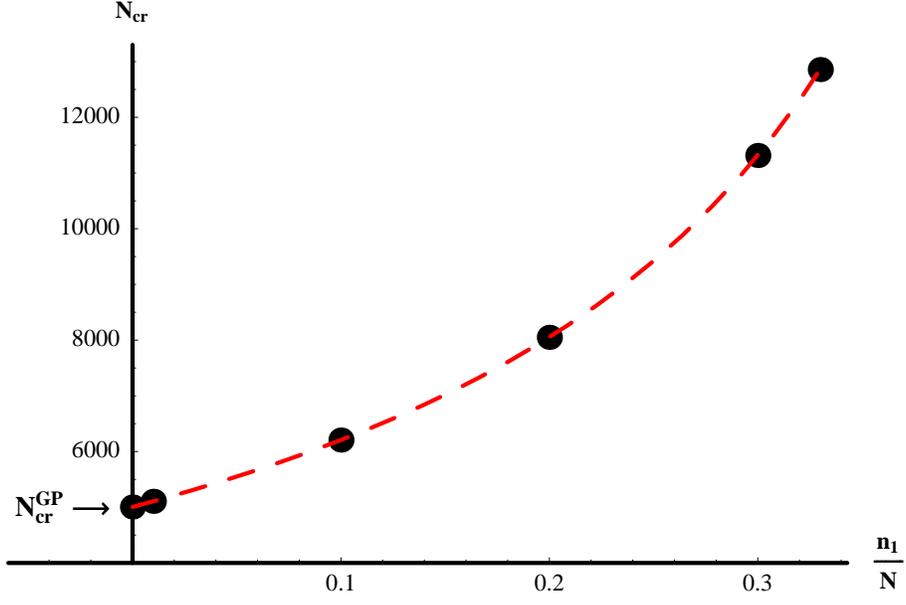}
\caption [kdv] {(Color online) The critical number $N_{\mathrm{cr}}$ obtained
with $E(\sigma_0,\sigma_1)$ of Eq.~(\ref{e6}) (black solid dots)
and with the analytic result of Eq.~(\ref{e7}) (red dashed curve)
as a function of the relative fragmentation $\frac{n_1}{N}$.
The quantities used are the same as in Fig.~\ref{f1}.
$N_{\mathrm{cr}}^{\mathrm{GP}}$ is indicated by an arrow. 
}
\label{f3}
\end{figure}

Although the dependence of $E$ on $\sigma_0$ and $\sigma_1$ is rather
simple, the analytical expression for $N_{\mathrm{cr}}$ is rather involved and not 
particularly informative.
For this reason we computed the critical value $N_{\mathrm{cr}}^{\sigma_0=\sigma_1}$
obtained from $E$ in (\ref{e6}) by setting $\sigma_0=\sigma_1$.
This can be done completely analogously to the calculation
of $N_{\mathrm{cr}}^{\mathrm{GP}}$ discussed above.
One readily obtains for $N \gg 1$
\beq\label{e7}
N_{\mathrm{cr}}^{\sigma_0=\sigma_1} =
N_{\mathrm{cr}}^{\mathrm{GP}} \cdot \frac{1+2\frac{n_1}{N}}{1-\frac{13}{4}\left(\frac{n_1}{N}\right)^2}
\eeq
which gives $\frac{60}{23} \approx 2.6$ as an upper limit for the ratio
$N_{\mathrm{cr}}^{\sigma_0=\sigma_1}/N_{\mathrm{cr}}^{\mathrm{GP}}$.
The correct result for this ratio is very similar to the simple estimate (\ref{e7})
as can be seen in Fig.~\ref{f3}.

Having demonstrated that fragmentation leads to metastable
states with $N$ well above $N_{\mathrm{cr}}^{\mathrm{GP}}$,
we would like to give an ultimate numerical proof of this statement.
To this end, we search fully numerically for a minimum of the energy functional (\ref{e4})
in the space of the orbitals $\varphi_0(r)$ and $\varphi_1(r)$.
These orbitals are thus determined self-consistently,
similarly as done in the case of the single orbital in GP theory.
Minimizing $E[\varphi_0,\varphi_1]$ leads to coupled
equations for the orbitals which are solved numerically on a grid self-consistently.
For details on the general multi-orbital equations and their numerical solution, 
we refer to the literature \cite{PLA03,PLA05,ALN,path,z0}.
In Fig.~\ref{e4} we show the solutions
$\varphi_0(r)$ and $\varphi_1(r)$ for $N=1.7N_{\mathrm{cr}}^{\mathrm{GP}}$,
and compare them with those obtained using the ``Gaussian model'' (\ref{e6}) above.
It is seen, in agreement with the model, that $\varphi_0$ and $\varphi_1$
simulate radial harmonics.
The observed deviations are due to the full self-consistency
of the numerical solution of the energy functional (\ref{e4}).

It is a rather very extensive task to explore the whole range of $\Lambda_0$,$N$
and relative fragmentation $\frac{n_1}{N}$
using the full optimization of the energy functional $E$ in (\ref{e4}).
We have,
therefore, performed calculations for several additional
examples only.
All the results obtained with this functional strongly support the
findings of the ``Gaussian model''.
In particular, we may conclude beyond doubt that fragmentation leads to metastable states of attractive condensates
with particle numbers much larger than possible for unfragmented ones.
In contrast to repulsive condensates where the fragmentation
tends to separate the fragments in space,
the fragmentation in attractive condensates is seen to be
``on-top'', i.e., the fragments prefer to be close to
each other and to live in the same part of space.

We would like to conclude with three brief remarks.
We have numerical indications for a further increase of $N_{\mathrm{cr}}$
when allowing for an even higher order of fragmentation,
i.e., by adding $d$-type ($l=2$) harmonics.
The increase of $N_{\mathrm{cr}}$ is probably only moderate.
We reiterate that the present findings are not restricted to spherical symmetry
and harmonic traps.
Fragmentation is expected to play a crucial role in the dynamics of attractive
condensates, in particular of atom-rich condensates.

\begin{acknowledgments}
Financial support by the DFG is gratefully acknowledged.
\end{acknowledgments}

\newpage

\begin{figure}[ht]
\includegraphics[width=12cm,angle=-0]{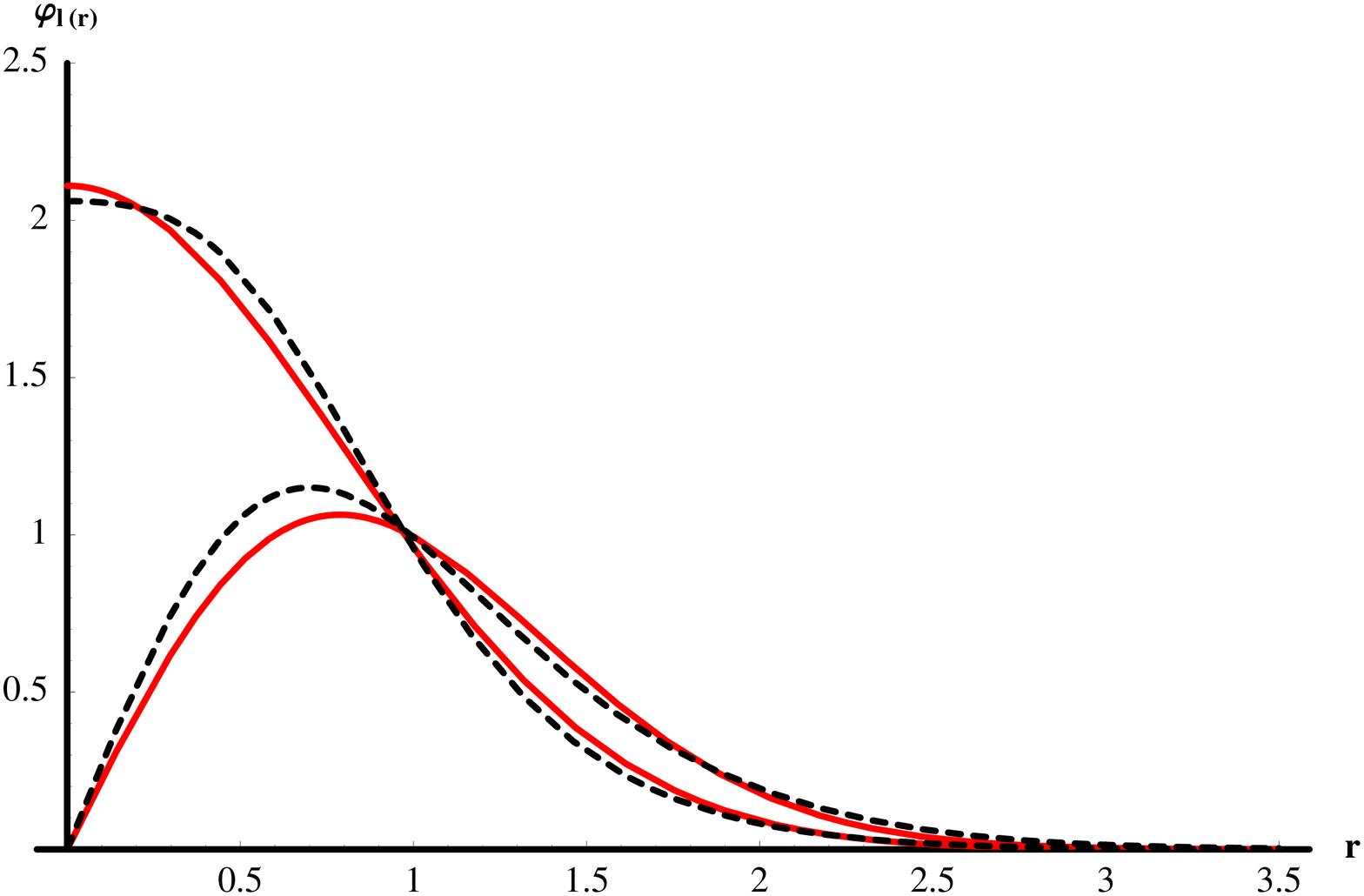}
\caption [kdv] {(Color online) The orbitals $\varphi_0$(r) and $\varphi_1(r)$ 
obtained fully numerically by minimizing the energy functional (\ref{e4})
for $N=1.7N_{\mathrm{cr}}^{\mathrm{GP}}$ (in black).
The results are compared with those of the ``Gaussian model''
using Eq.~(\ref{e6}) (in red).
Here, $\sigma_0=0.797$ and $\sigma_1=0.788$ at the
minimum of the energy in Eq.~(\ref{e6}).
The relative fragmentation is $\frac{n_1}{N}=0.25$.
The other quantities used are the same as in Fig.~\ref{f1}.
}
\label{f4}
\end{figure}


\begin{thebibliography}{99}

\bibitem{e1} C. C. Bradley {\it et al.}, Phys. Rev. Lett. {\bf 75}, 1687 (1995).

\bibitem{e2} S. L. Cornish {\it et al.},
 Phys. Rev. Lett. {\bf 85}, 1795 (2000).

\bibitem{e3} E. A. Donely {\it et al.},
 Nature (London) {\bf 412}, 295 (2001). 

\bibitem{e4} L. Khaykovich {\it et al.},
 Science {\bf 296}, 1290 (2001). 

\bibitem{e5} T. Weber {\it et al.},
 Science {\bf 299}, 232 (2003).

\bibitem{e6} S. L. Cornish {\it et al.},
 Phys. Rev. Lett. {\bf 96}, 170401 (2006). 

\bibitem{t1} P. A. Ruprecht {\it et al.},
 Phys. Rev. A {\bf 51}, 4704 (1995).

\bibitem{ga1} A. L. Fetter, cond-mat/9510037.

\bibitem{t2} F. Dalfovo and S. Stringari,
 Phys. Rev. A {\bf 53}, 2477 (1996).

\bibitem{ct1} V. M. P\'erez-Garc\'ia {\it et al.},
 Phys. Rev. Lett. {\bf 77}, 5320 (1996); 
 Phys. Rev. A {\bf 56}, 1424 (1997).

\bibitem{ga2} H. T. C. Stoof, J. Stat. Phys. {\bf 87}, 1353 (1997).

\bibitem{t4} Yu. Kagan {\it et al.},
 Phys. Rev. Lett. {\bf 81}, 933 (1998).

\bibitem{ga3} M. Ueda and A. J. Leggett,
 Phys. Rev. Lett. {\bf 80}, 1576 (1998).

\bibitem{t5} R. A. Duine and H. T. C. Stoof,
 Phys. Rev. Lett. {\bf 86}, 2204 (2001).

\bibitem{ct2} H. Saito and M. Ueda,
 Phys. Rev. A {\bf 63}, 043601 (2001).

\bibitem{t6} L. Santos and G. V. Shlyapnikov,
 Phys. Rev. A {\bf 66}, 011602(R) (2002). 

\bibitem{t7} L. D. Carr and J. Brand,
 Phys. Rev. Lett. {\bf 92}, 040401 (2004).

\bibitem{t8} S. W\"uster {\it et al.},
 Phys. Rev. A {\bf 71}, 033604 (2005).

\bibitem{t9} S. K. Adhikari,
 Phys. Rev. A {\bf 71}, 053603 (2005).

\bibitem{t10} A. D. Martin {\it et al.},
 Phys. Rev. Lett. {\bf 98}, 020402 (2007). 

\bibitem{Pethick} C. J. Pethick and H. Smith, {\it Bose-Einstein Condensation in Dilute Gases} (Cambridge
 Univ. Press, Cambridge, 2002).

\bibitem{Pit_book} L. Pitaevskii and S. Stringari, {\it Bose-Einstein Condensation}
(Oxford Univ. Press, Oxford, 2003).

\bibitem{PLA03} L. S. Cederbaum and A. I. Streltsov, 
 Phys. Lett. A {\bf 318}, 564 (2003).

\bibitem{PLA05} O. E. Alon {\it et al.},
 Phys. Lett. A {\bf 347}, 88 (2005).

\bibitem{frag0} P. Nozi\'eres, in {\it Bose-Einstein Condensation}, 
 eds. A. Griffin {\it et al.}, (Cambridge Univ. Press, Cambridge, 1996).

\bibitem{Sipe} R. W. Spekkens and J. E. Sipe, 
 Phys. Rev. A {\bf 59}, 3868 (1999).

\bibitem{ALN} A. I. Streltsov {\it et al.},
 Phys. Rev. A {\bf 70}, 053607 (2004).

\bibitem{path} O. E. Alon and L. S. Cederbaum, 
 Phys. Rev. Lett. {\bf 95}, 140402 (2005).

\bibitem{Erich} E. J. Mueller {\it et al.},
 Phys. Rev. A {\bf 74}, 033612 (2006).

\bibitem{Girardeau} M. Girardeau, 
 J. Math. Phys. (N.Y.) {\bf 1}, 516 (1960).

\bibitem{Fisher} M. P. A. Fisher {\it et al.},
 Phys. Rev. B {\bf 40}, 546 (1989).

\bibitem{Zoller} D. Jaksch {\it et al.},
 Phys. Rev. Lett. {\bf 81}, 3108 (1998).

\bibitem{Bloch} M. Greiner {\it et al.},
 Nature (London) {\bf 415}, 39 (2002).

\bibitem{z0} O. E. Alon {\it et al.},
 Phys. Rev. Lett. {\bf 95}, 030405 (2005).

\bibitem{Ueda97} M. Ueda, A. J. Leggett,
 Phys. Rev. Lett. {\bf 83}, 1489 (1999).

\bibitem{Ueda03} R. Kanamoto {\it et al.},
 Phys. Rev. A {\bf 67}, 013608 (2003).

\bibitem{EPL} O. E. Alon {\it et al.},
 Europhys. Lett. {\bf 67}, 8 (2004).
 
\end{thebibliography}
\end{document}